\documentclass[aps,twocolumn,floatfix,showpacs]{revtex4}
\usepackage{graphicx}

\begin{document}

\title{Dynamical Response of Networks under External Perturbations:
Exact Results}

\author{David D. Chinellato$^1$, Marcus A.M. de Aguiar$^{1,2}$,
Irving R. Epstein$^{2,3}$, Dan Braha$^{2,4}$ and Yaneer Bar-Yam$^2$}

\affiliation{$^1$Instituto de F\'{\i}sica `Gleb Wataghin',
Universidade Estadual de Campinas, Unicamp 13083-970, Campinas, SP,
Brasil\\ $^2$New England Complex Systems Institute, Cambridge,
Massachusetts 02138 \\$^3$Department of Chemistry, MS015, Brandeis
University, Waltham, Massachusetts 02454, USA \\ $^4$University of
Massachusetts, Dartmouth, Massachusetts 02747}

\begin{abstract}

We introduce and solve a general model of dynamic response under
external perturbations. This model captures a wide range of systems
out of equilibrium including Ising models of physical systems,
social opinions, and population genetics. The distribution of
states under perturbation and relaxation process reflects two
regimes --- one driven by the external perturbation, and one driven
by internal ordering. These regimes parallel the disordered and
ordered regimes of equilibrium physical systems driven by thermal
perturbations but here are shown to be relevant for non-thermal and
non-equilibrium external influences on complex biological and
social systems. We extend our results to a wide range of network
topologies by introducing an effective strength of external
perturbation by analytic mean-field approximation. Simulations show
this generalization is remarkably accurate for many topologies of
current interest in describing real systems.

\end{abstract}

\pacs{89.75.-k,05.50.+q,05.45.Xt}

\maketitle

Networks have become a standard model for a wealth of complex
systems, from physics to social sciences to biology
\cite{bararev,hwa06}. A large body of work has investigated
topological properties \cite{bararev,baryam2004,alb,cohen}. The
{\it raison d'\^etre}, though, of complex network studies is to
understand the relationship between structure and dynamics - from
disease spreading and social influence
\cite{ves01,pecora02,motter03,pac04,laguna} to search\cite{guime}.
Yet, dynamic response of networks under external perturbations has
been less thoroughly investigated \cite{baryam2004,xing}. In this
paper we consider a simple dynamical process as a general framework
for the dynamic response of a network to an external environment.
The environment is initially treated as part of the network and
then generalized as an external system.

We obtain complete and exact results for the simplest case of fully
connected networks and find a nontrivial dynamic behavior that can
be divided into two regimes. For large perturbations the
environmental influence extends into the system with a distribution
which, in the thermodynamic limit, becomes a Gaussian around a
value that reflects a balance between the external perturbations
driving the system in different directions. For small perturbations
the distribution of states has peaks at the two ordered states.
Order arises from interactions within the system, and power law
tails result from the external perturbation away from these ordered
states. The boundary between these regimes is characterized by a
uniform distribution where all states are equally likely. The time
scale of equilibration is small for large perturbations and
diverges inversely as the strength of the perturbation for small
perturbations. This characterizes the switching time behavior of
the two ordered states. We generalize the exact results to networks
of different topologies using a mean field treatment. Simulations
show that this generalization, which involves renormalizing the
constants in the distributions, is very accurate. Our results
reveal and generalize key features of relaxation and dynamic
response of models of a wide range of physical systems in the Ising
universality class, electoral and contagion models of social
systems, and the Wright-Fisher model of evolution in population
biology.

Specifically, we consider networks with $N+N_0+N_1$ nodes. Each
node has an internal state which can take only the values $0$ or
$1$. We let the $N_0$ nodes be frozen in state 0, and $N_1$ in
state 1, and the remaining nodes change by adopting the state of a
connected node. At each time step a random free node is selected;
with probability $1-p$ the node copies the state of one of its
connected neighbors, and with probability $p$ the state remains
unchanged. The frozen nodes can be interpreted as external
perturbations to the subnetwork of free nodes. Analytically
extending $N_0$ and $N_1$ to be smaller than 1 enables modeling the
case of weak coupling. This model generalizes our previous efforts
to derive exact results of network dynamics \cite{aguiar05} (see
also\cite{zhoulipowsky2007}).

This system is similar to the Ising model, where $N_1+N_0$ by
explicitly representing the impact of thermal perturbations play the
roles of the temperature $T$, and $N_1-N_0$ acts as an external
magnetic field $h$. Our dynamics are equivalent to Glauber dynamics
\cite{glauber} for weak fields and high temperatures, where the
Ising model parameters are $J/kT \rightarrow 1/(z+N_0+N_1)$ and $h/J
\rightarrow (N_1-N_0)$, where $z$ is the number of nearest neighbors
and $J$ the nearest-neighbor interaction strength. For low
temperatures our model is an alternative dynamics that also captures
the key kinetic properties of this system. Relevant network
structures include crystalline 3-D lattices and random networks for
amorphous spin-glasses; fully connected networks correspond to long
range interactions or the mean field approximation. Despite the
relevance to the extensively studied Ising model, we are not aware
of any other exact solution of the response dynamics of a fully
connected system or explicit representation of thermal or other
perturbation for dynamic response. Specific results are available
only for zero temperature dynamics in one-dimensional or mean field
systems. \cite{pradosbrey2001,spasojevic2006}

Our system can also model an election with two candidates
\cite{vilone,redner} where some of the voters have a fixed opinion
while the rest change their intention according to the opinion of
others. Another application is to epidemics that spread upon
contact between infected nodes (e.g., individuals or computers).
Finally, the model can represent an evolving population of sexually
reproducing (haploid) organisms where the internal state represents
one of two alleles of a gene \cite{ewens}. Taking $p=1/2$, the
update of a node mimics the mating of two individuals, with one
parent being replaced by the offspring, which can receive the
allele of either the mother or the father with $50\%$ probability.
Since a free node can also copy the state of a frozen node, the
ratios $N_0/(N+N_0+N_1-1)$ and $N_1/(N+N_0+N_1-1)$ give the
mutation rates.

For a fully connected network the nodes are indistinguishable and
the state of the network is fully specified by the number of nodes
with internal state 1 \cite{aguiar05}. Therefore, there are only
$N+1$ global states, which we denote $\sigma_k$, $k=0,1,...,N$. The
state $\sigma_k$ has $k$ free nodes in state $1$ and $N-k$ free
nodes in state $0$. If $P_t(m)$ is the probability of finding the
network in the state $\sigma_m$ at the time $t$, then $P_{t+1}(m)$
can depend only on $P_t(m)$, $P_t(m+1)$ and $P_t(m-1)$. The
probabilities $P_t(m)$ define a vector of $N+1$ components ${\bf
P}_t$. In terms of ${\bf P}_t$ the dynamics is described by the
equation
\begin{displaymath}
{\bf P}_{t+1} = {\bf U} {\bf P}_t \equiv \left( {\bf 1} -
\frac{(1-p)}{N(N+N_0+N_1-1)} {\bf A}\right) {\bf P}_t \label{timev}
\end{displaymath}
where the {\it evolution matrix} ${\bf U}$, and also the auxiliary
matrix ${\bf A}$, is tri-diagonal. The non-zero elements of ${\bf
A}$ are independent of $p$ and are given by
\begin{displaymath}
\begin{array}{l}
A_{m,m} = 2m(N-m) + N_1(N-m) + N_0 m \\
A_{m,m+1} = -(m+1)(N+N_0-m-1) \\
A_{m,m-1} = -(N-m+1)(N_1+m-1).
\end{array}
\end{displaymath}
The transition probability from state $\sigma_M$ to $\sigma_L$ after
a time $t$ can be written as
\begin{equation}
P(L,t;M,0) = \sum_{r=0}^N \frac{1}{\Gamma_r} b_{rM} a_{rL}
\lambda_r^t \;. \label{problm}
\end{equation}
where $a_{rL}$ and $b_{rM}$ are the components of the right and
left r-th eigenvectors of the evolution matrix, ${\bf a}_r$ and
${\bf b}_r$, with $\Gamma_r = {\bf b}_r \cdot {\bf a}_r$. Thus, the
dynamical problem has been reduced to finding the right and left
eigenvectors and the eigenvalues of ${\bf A}$.

It is easy to check by inspection of small matrices that the
eigenvalues $\mu_r$ of ${\bf A}$ are given by
\begin{displaymath}
\mu_r = r(r-1+N_0+N_1).
\end{displaymath}
This implies $0 \leq p \leq \lambda_r \leq 1$, where $\lambda_r$
are the eigenvalues of ${\bf U}$. Because of Eq.(\ref{problm}), the
unit eigenvalues completely determine the asymptotic behavior of
the system.

The eigensystem ${\bf A} {\bf a}_r = \mu_r {\bf a}_r$ leads to the
following recursion relation for the coefficients $a_{rm}$
\begin{equation}
\sum_{j=m-1}^{m+1} A_{mj} \, a_{rj} = \mu_r \, a_{rm}
\end{equation}
with $a_{r,N+1}=a_{r,-1}\equiv 0$. To solve this equation we
multiply the whole expression by $x^m$, sum over $m$ and define the
generating function $p_r(x) = \sum_{m=0}^N a_{rm} x^m$. The
recursion relation then yields the following differential equation
for $p_r$
\begin{equation}
\begin{array}{l}
x(1-x) p_r'' + [(1-N-N_0) - (1+N_1-N)x]p_r' + \\
~[N N_1 -\mu_r/(1-x)] p_r = 0. \label{genr}
\end{array}
\end{equation}

To understand the asymptotic behavior of the system ($\mu_r=0$) we have
to consider two cases:\\

(a) If $N_0 = N_1 = 0$ then $\mu_r=0$ leads to $r=0$ or $r=1$
\cite{aguiar05}. In this case the differential equation simplifies
to $xp_r''+(1-N)p_r'=0$, whose two independent solutions are
$p_0(x)=1$ and $p_1(x)=x^N$, corresponding to the all--nodes--0 or
all--nodes--1 states respectively. \\

(b) If $N_0, N_1 \neq 0$ then $\mu_r=0$ implies $r=0$. In this case
equation (\ref{genr}) is that of a hypergeometric function $F$ and
we find $p_0(x) = F(-N,N_1,1-N-N_0,x)$, which is a finite
polynomial with known coefficients $a_{0m}$. Normalizing this
eigenvector, we obtain the probability of finding the network in
state $\sigma_m$ at large times:
\begin{equation}
\rho(m) =  {\cal A} ~ \frac{(N_1+m-1)!~(N + N_0 - m -
1)!}{(N-m)!~m!} \label{probn}
\end{equation}
where ${\cal A}={\cal A}(N,N_0,N_1)$ is a normalization. Because of
the frozen nodes, the dynamics will never stabilize in any state,
but will always move from one state to another, with mean
occupation number $\bar{m} = N N_1/(N_0+N_1)$. The surprising
feature of this solution is that for $N_0=N_1=1$ we obtain
$\rho(m)=1/(N+1)$, for all values of $N$. Thus all macroscopic
states are equally likely and the system executes a random walk
through the state space.

The dynamics at long times is dominated by the second largest
eigenvector, with eigenvalue $\lambda_1$. For large networks
$\lambda_1^t \approx e^{-t/\tau}$ where
\begin{equation}
\tau=\frac{N(N+N_0+N_1-1)}{(1-p)(N_0+N_1)}. \label{tau}
\end{equation}

We obtain a complete description of the dynamics by deriving all
eigenvectors with $\mu_r \neq 0$. The differential equation for
$p_r(x)$ yields
\begin{equation}
p_r(x) = {\textstyle \frac{
F(1-r-N_0,1-r-N-N_0-N_1,1-N-N_0,x)}{(1-x)^{r-1+N_0+N_1}}} \, .
\label{eqp}
\end{equation}
Expanding the numerator and denominator in Taylor series gives the
coefficients $a_{rm}$. Although they can easily be written down
explicitly, we  do not do so here. Similarly, defining the
generating function $q_r(x) = \sum_{m=1-N_1}^{N+N_0-1} b_{rm} x^m$
we obtain a differential equation for $q_r$ whose solution is
\begin{equation}
\begin{array}{l}
q_r(x) = {\textstyle \frac{x^{1-N_1}~
F(1-r-N_1,1-r-N-N_0-N_1,1-N-N_1,x)}{(1-x)^{r+1}} } . \label{eqq}
\end{array}
\end{equation}
If $N_0 = N_1 = 0$ this solution is not valid for $r=0$ or $r=1$,
since the matrix ${\bf A}^T$ becomes singular. In this case the two
left eigenvectors are given by $b_{0,m} = 1$ and $b_{1,m} = N -2m$.
For other cases the solution is obtained from the power series
expansion of $q_r(x)$. Equations (\ref{eqp}) and (\ref{eqq})
complete the solution of the problem.

In the thermodynamic limit $N\rightarrow \infty$ we can define
continuous variables $x=m/N$, $n_0=N_0/N$ and $n_1=N_1/N$ and
approximate the asymptotic distribution by a Gaussian $\rho(x) =
\rho_0 \exp{[-(x-x_0)^2/2\delta^2]}$ with $x_0=n_1/(n_0+n_1)$,
$\rho_0=1/\sqrt{2\pi\delta^2}$ and
\begin{equation}
\delta = \left[\frac{n_0 n_1(1+n_0+n_1)}{N(n_0+n_1)^3}\right]^{1/2}.
\end{equation}
In the limit where $n_0,~n_1 >> 1$ the width depends only on the
ratio $\alpha=n_0/n_1$ and is given by $\sqrt{\alpha/N}/(1+\alpha)$.
\begin{figure}
   \includegraphics[clip=true,width=6cm,angle=-90]{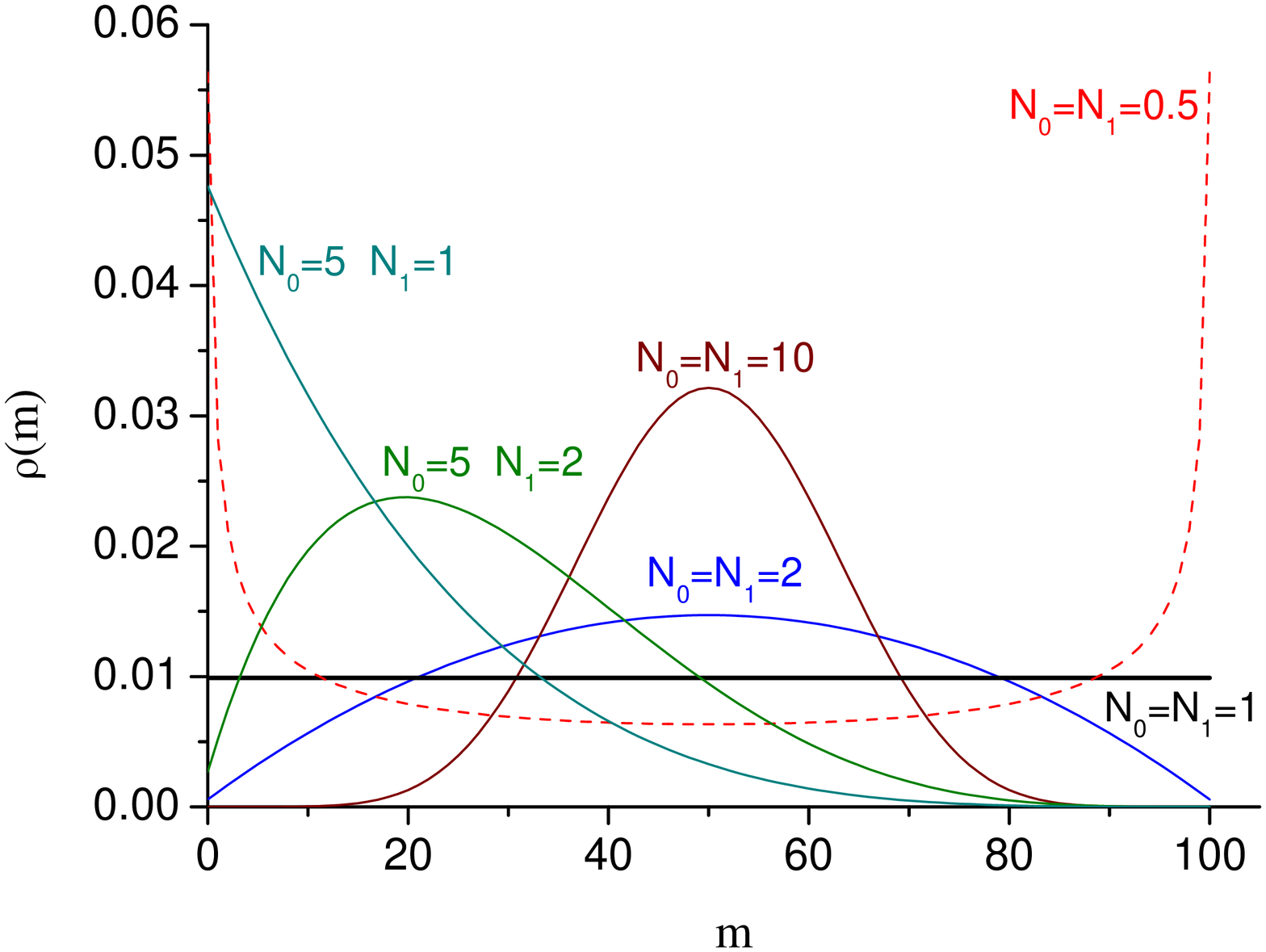}
   \caption{(color online) Asymptotic probability distribution for a
   network with $N=100$ and several values of $N_0$ and $N_1$.}
\end{figure}

The problem we just solved can be generalized to treat an external
reservoir weakly coupled to the network of $N$ nodes. We note that
the differential equations for the generating functions $p_r(x)$
and $q_r(x)$ remain well defined for real $N_0$ and $N_1$. The
solutions for the generating functions remain the same, except that
factorials must be replaced by gamma functions. Since
$N_0/(N+N_0+N_1-1)$ and $N_1/(N+N_0+N_1-1)$ represent the
probabilities that a free node copies one of the frozen nodes,
small values of $N_0$ and $N_1$ can be interpreted as representing
a weak connection between the free nodes and an external system
containing the frozen nodes. The external system can be thought of
as a reservoir that affects the network but is not affected by it.
Alternatively, we can suppose that there is a single node fixed at
0 that is on for only a fraction $N_0$ of the time and off for the
fraction $1 - N_0$, and similarly for a single node fixed at 1.

Figure 1 shows examples of the distribution $\rho(m)$ for a network
with $N=100$ and various values of $N_0$ and $N_1$. Numerical
simulations displaying similar results are described in
\cite{boccara}.

Figure 2 shows an example of the time evolution of the probability
density for a fully connected network compared to numerical
simulations. The evolution from the initial to the asymptotic
time-independent distribution is the analog of an equilibration
process promoted by the external system.

\begin{figure}
   \includegraphics[clip=true,width=2.925cm,angle=-90]{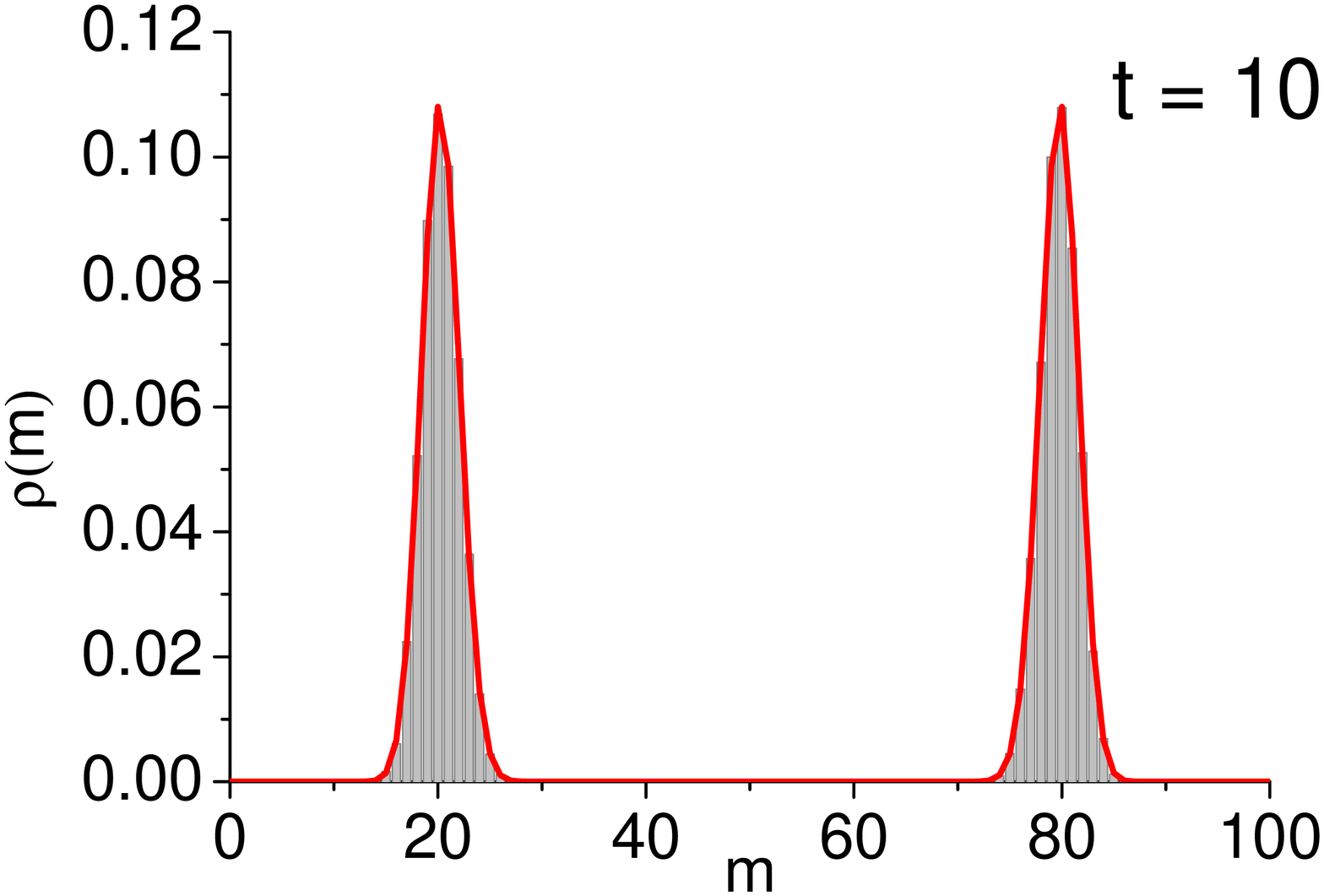}
   \includegraphics[clip=true,width=2.925cm,angle=-90]{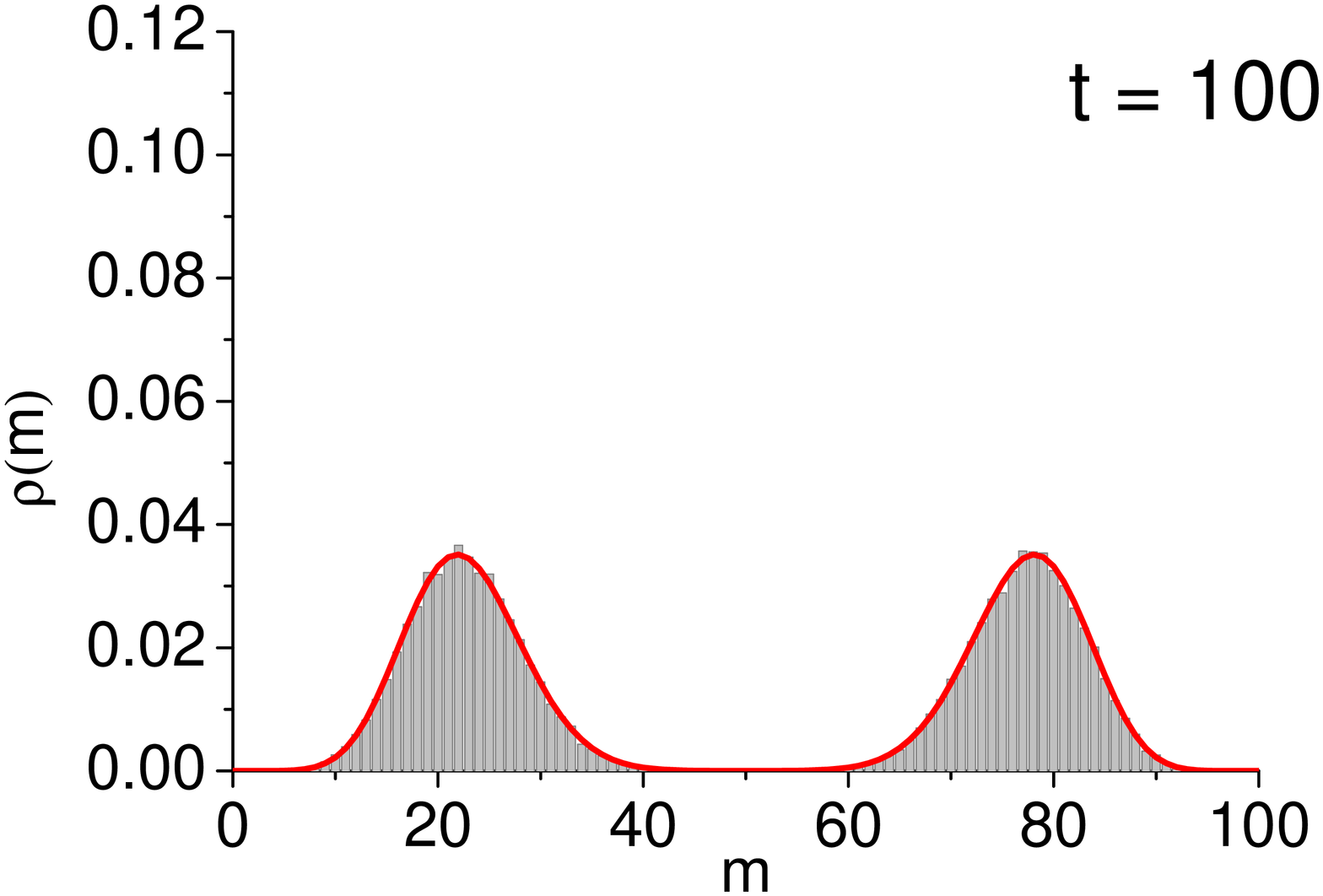}
   \includegraphics[clip=true,width=2.925cm,angle=-90]{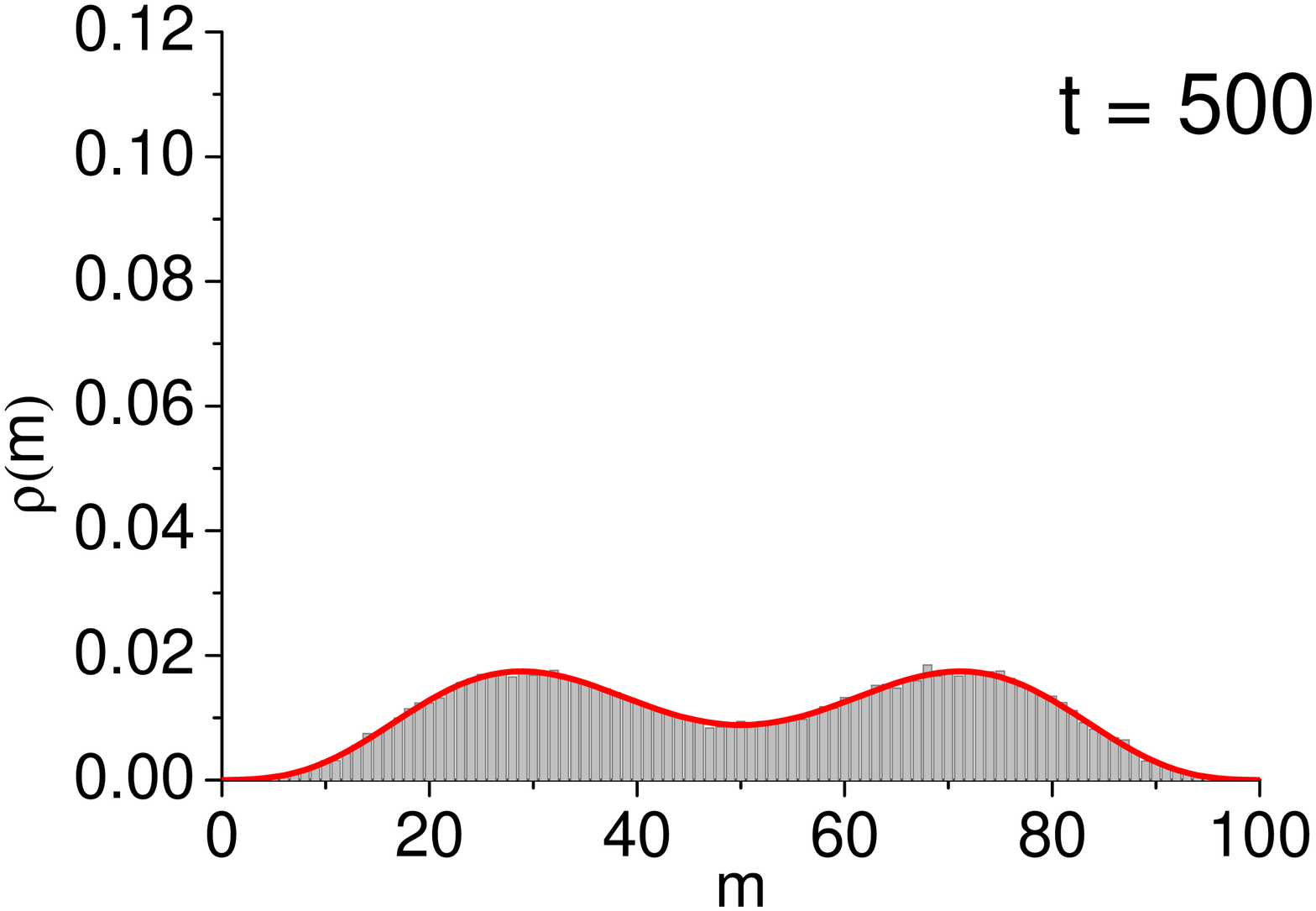}
   \includegraphics[clip=true,width=2.925cm,angle=-90]{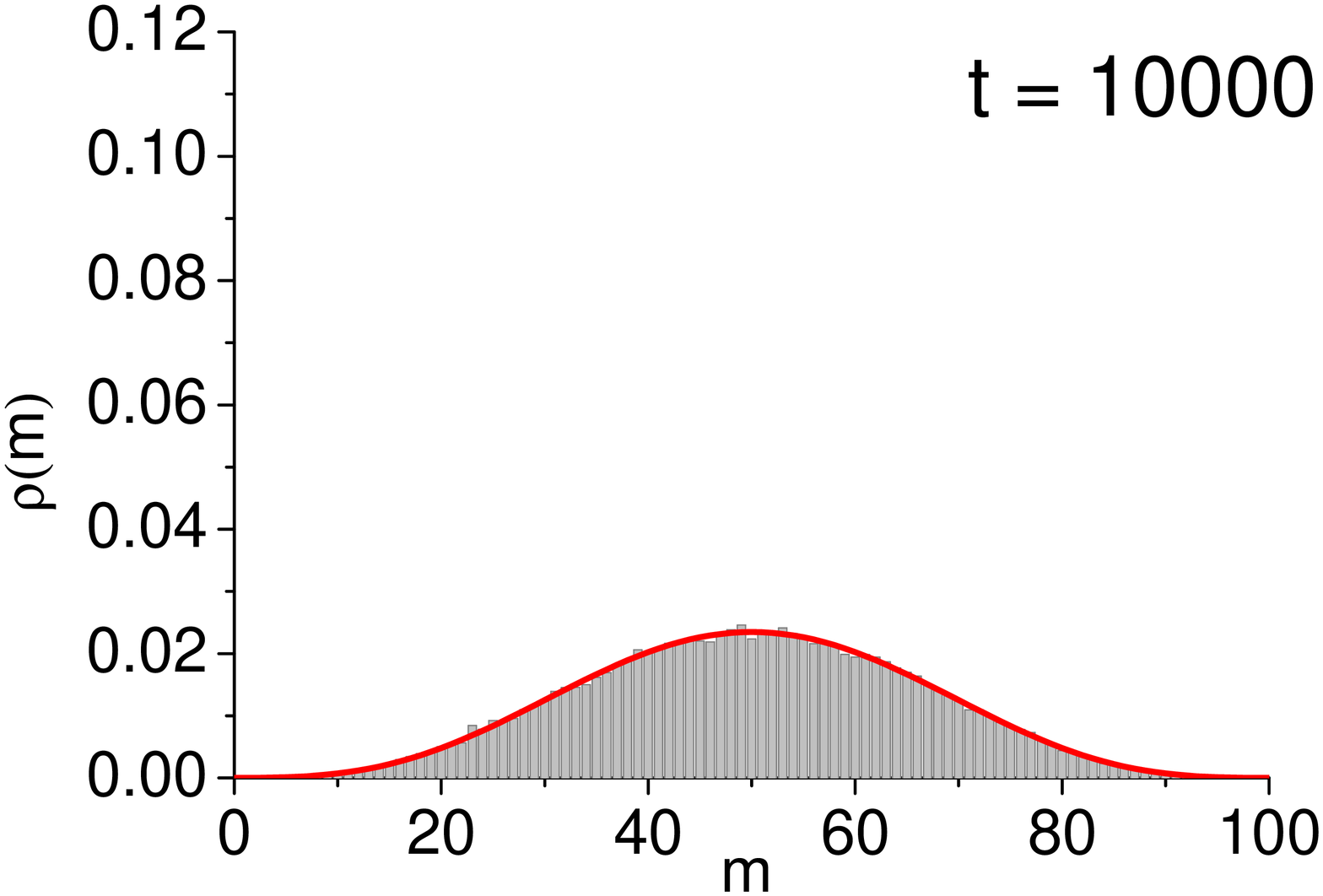}
   \caption{(color online) Time evolution of the probability distribution $P_t$
   for a network with $N=100$ and $N_0=N_1=5$. The initial state is
   $P_{0i}=0.5(\delta_{i,20}+\delta_{i,80})$. The histograms show the
   average over 50,000 actual realizations of the dynamics and the solid
   (red) line shows the analytical result.}
\end{figure}

For small values of $N_0$ and $N_1$ ($<< 1/\ln{N}$), we can obtain a
simplified expression for $\rho(m)$:
\begin{equation}
\rho(m) \approx \frac{N_1 N_0}{N_0+N_1}~\left[
\frac{1-N_1\ln{N}}{m^{1-N_1}} + \frac{1-N_0\ln{N}}{(N-m)^{1-N_0}}
\right].
\end{equation}
Thus $\rho(m)$ displays a power law behavior on both ends of the
curve: $1/m$ for $m$ close to 0 and $1/(N-m)$ for $m$ close to $N$
(see, for instance, the curve with $N_0=N_1=0.5$ in Fig. 1). Since
the relaxation time $\tau$ is proportional to $1/(N_0+N_1)$, the
equilibration process becomes very slow in this limit.

\begin{figure}
   \includegraphics[clip=true,width=2.925cm,angle=-90]{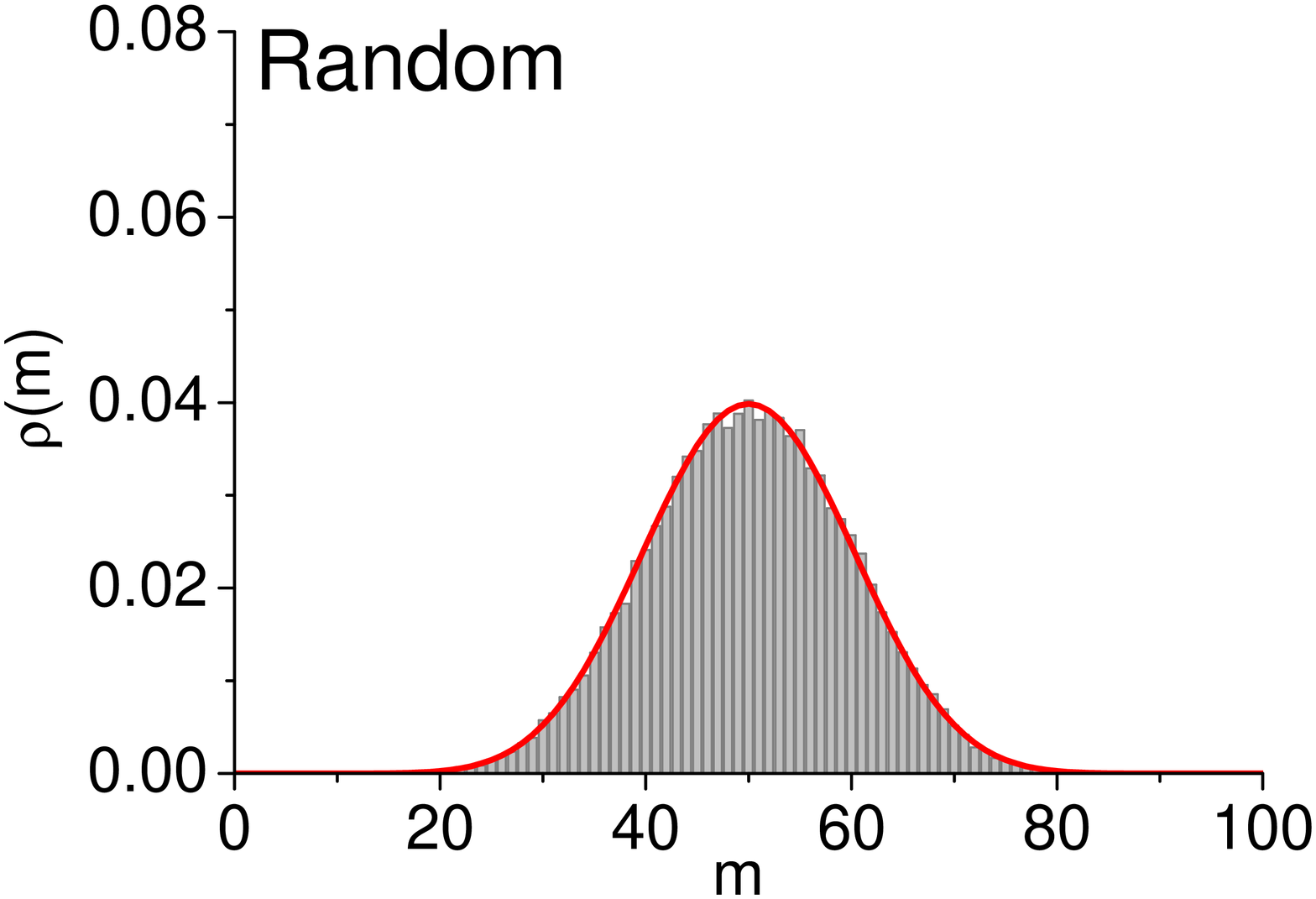}
   \includegraphics[clip=true,width=2.925cm,angle=-90]{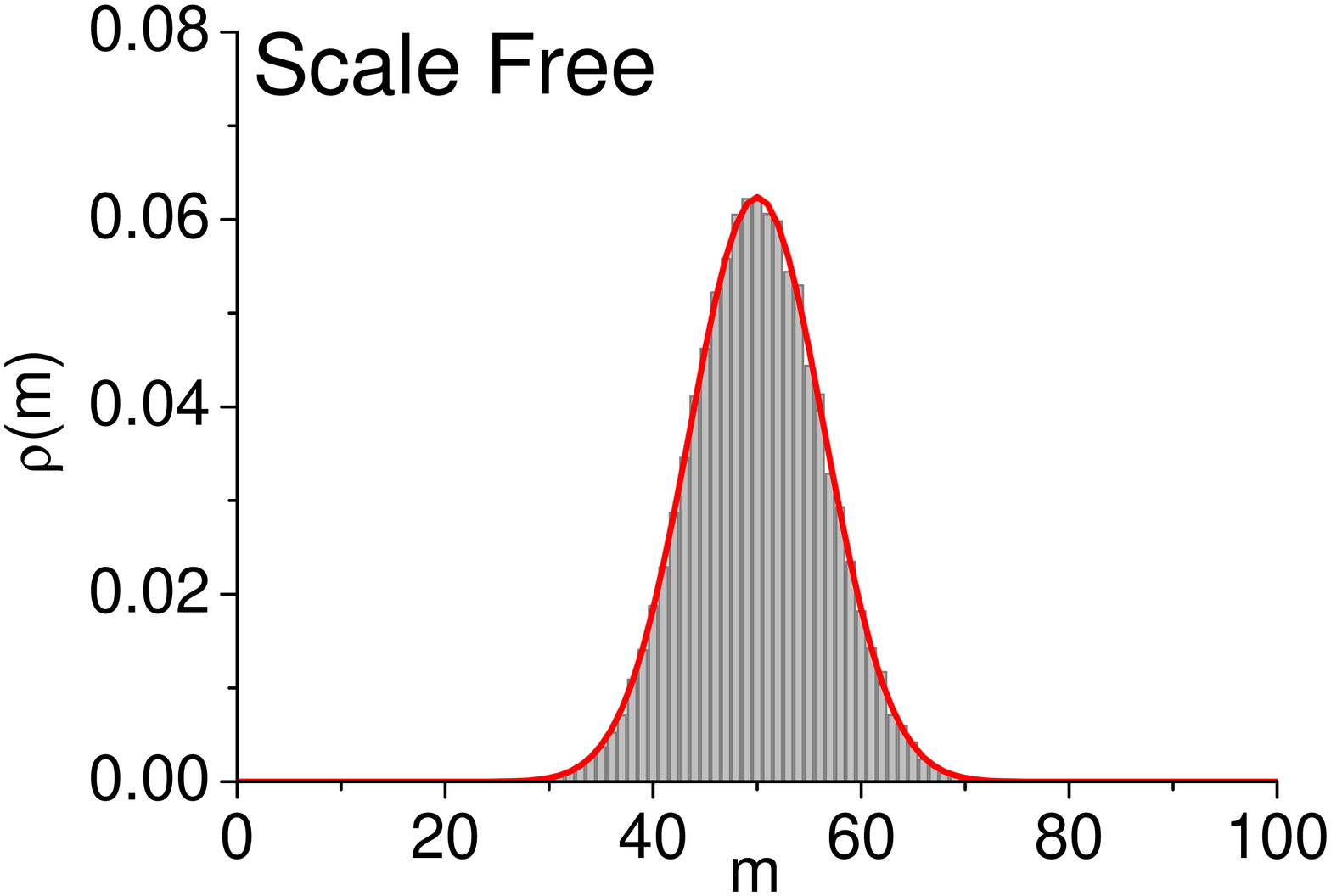}
   \includegraphics[clip=true,width=2.925cm,angle=-90]{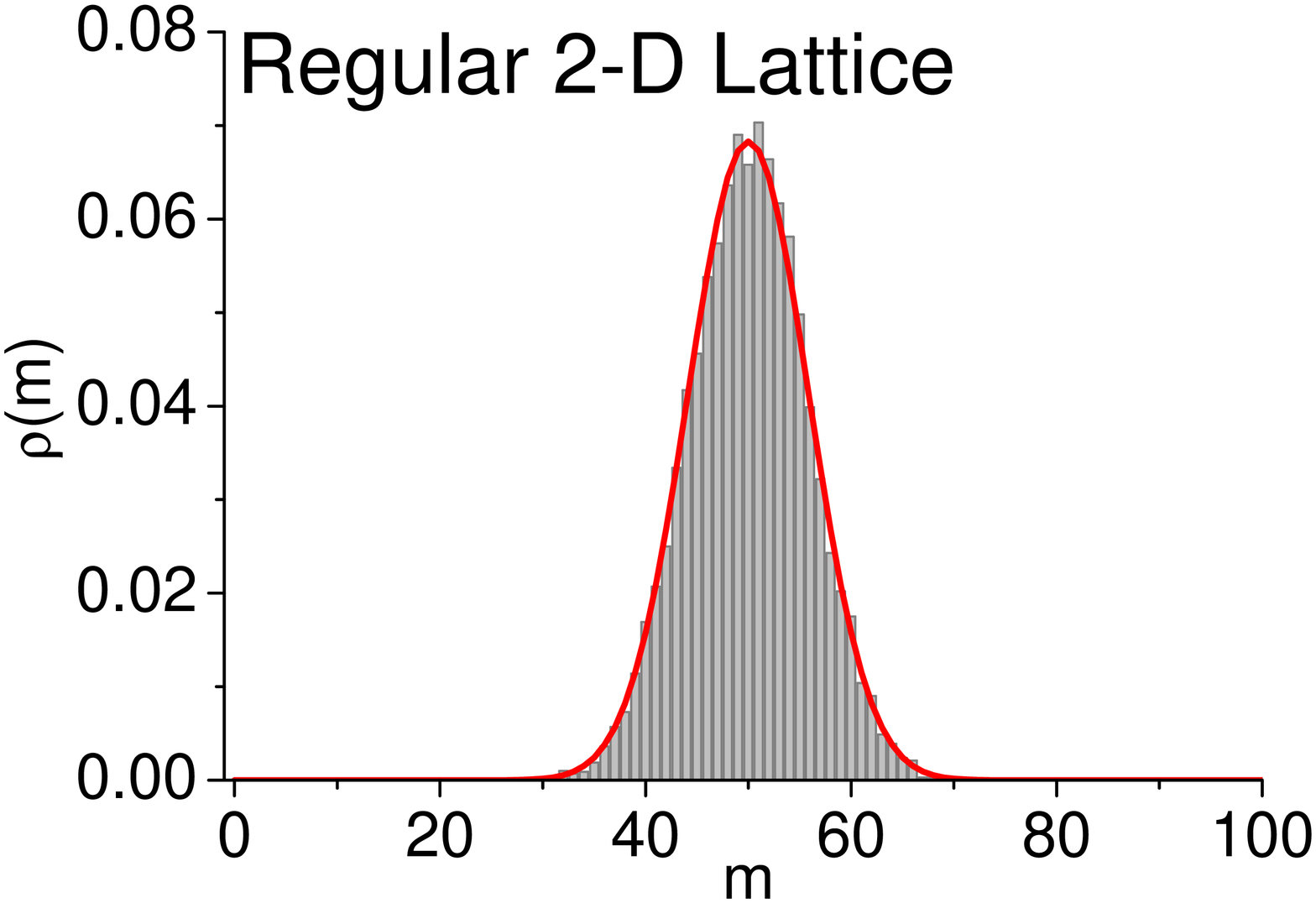}
   \includegraphics[clip=true,width=2.925cm,angle=-90]{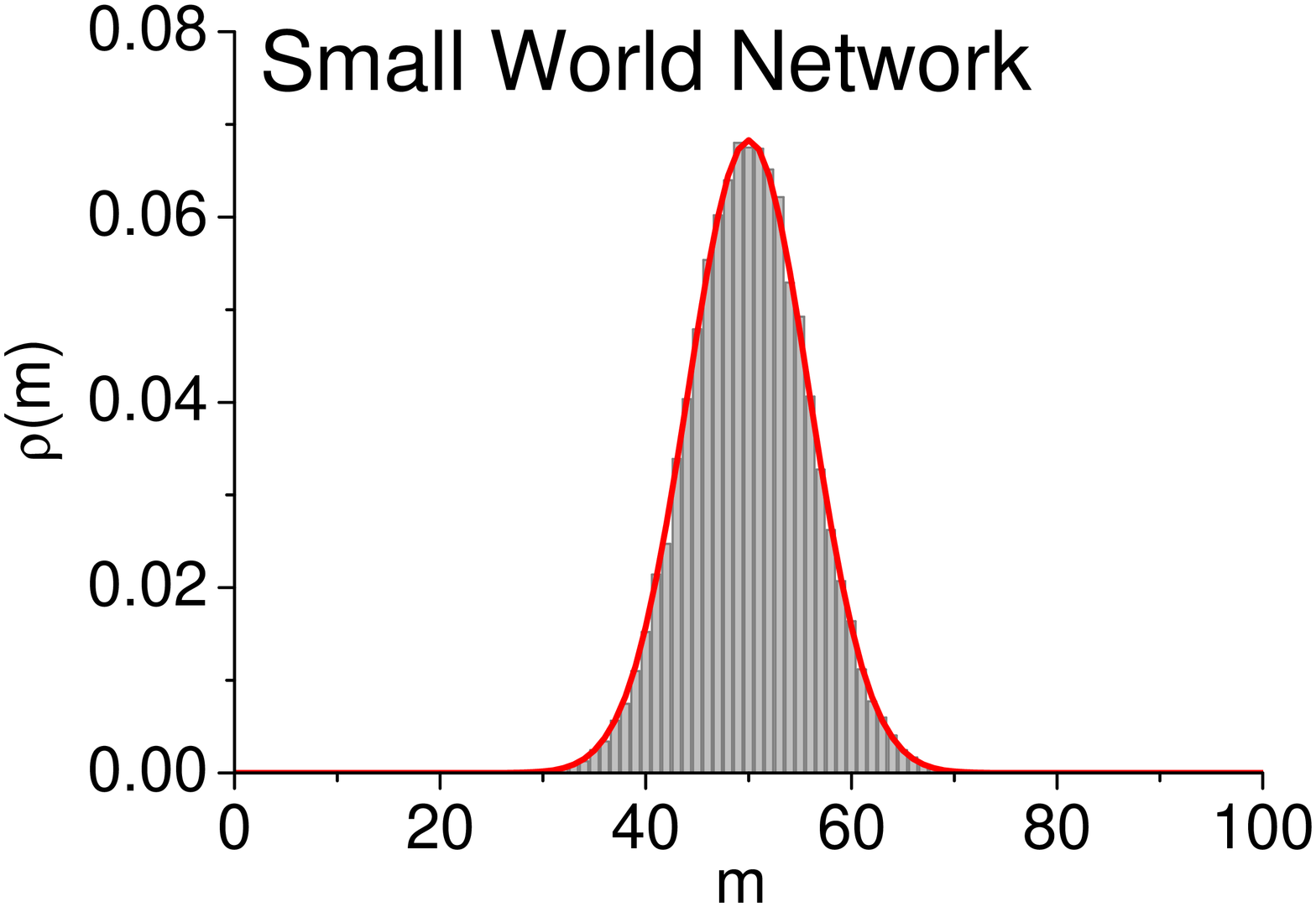}
   \caption{(color online) Asymptotic probability distribution for networks with
   different topologies. In all cases $N=100$, $N_0=N_1=5$, $t=10,000$, and
   the number of realizations is $50,000$. The theoretical (red) curve is drawn with
   effective numbers of frozen nodes $N_{0ef}=f N_0$ and $N_{1ef}=f N_1$:
   (a) random network $N_{0ef}=N_{1ef}=17$; (b) scale-free $N_{0ef}=N_{1ef}=82$;
   (c) regular 2-D lattice $N_{0ef}=N_{1ef}=140$;
   (d) small world network $N_{0ef}=N_{1ef}=140$.}
\end{figure}

For networks with different topologies the effect of the frozen
nodes is amplified. To see this we note that the probability that a
free node copies a frozen node is
$P_i=(N_{0}+N_{1})/(N_{0}+N_{1}+k_i)$ where $k_i$ is the degree of
the node. For fully connected networks $k_i=N-1$ and $P_i\equiv
P_{FC}$. For general networks an average value $P_{av}$ can be
calculated by replacing $k_i$ by the average degree $k_{av}$. We
can then define effective numbers of frozen nodes, $N_{0ef}$ and
$N_{1ef}$, as being the values of $N_{0}$ and $N_{1}$ in $P_{FC}$
for which $P_{av} \equiv P_{FC}$. This leads to
\begin{equation}
N_{0ef} = f N_{0}, \qquad \qquad N_{1ef} = f N_{1}
\label{eq:rescaling}
\end{equation}
where $f=(N-1)/k_{av}$. Corrections involving higher moments can be
obtained by integrating $P_i$ with the degree distribution and
expanding around $k_{av}$.

Figure 3 shows examples of the equilibrium distribution for four
different networks with $N=100$ and $N_0=N_1=5$. Panel (a) shows
 a random network with connection probability $0.3$
($N_{av}=30$, $f=3.3$). The theoretical result was obtained with
Eq.~(\ref{probn}) with $N_{0ef}=N_{1ef}=17$. For a scale-free
network (panel (b)) grown from an initial cluster of 6 nodes adding
nodes with 3 connections each following the preferential attachment
rule \cite{bararev}, $f=99/6$ and the effective values of $N_0$ and
$N_1$ are approximately 82. Panel (c) shows the probability
distribution for a 2-D regular lattice with $10 \times 10$ nodes
for which $f=99/3.6\approx 28$. Finally, panel (d) shows a small
world version of the regular lattice \cite{bararev}, where 30
connections were randomly re-connected, creating shortcuts between
otherwise distant nodes. These results show that the mean field
generalization is accurate for many network topologies. Still,
extreme cases such as a star network should be different and this
is confirmed by simulations and preliminary analytic results. The
relaxation time (\ref{tau}), in units of network size and for
$p=0$, becomes $\tau/N=(k_{av}+N_0+N_1)/(N_0+N_1)$. It increases
linearly with $N$ for fully connected or random networks, but is
independent of $N$ for regular and scale-free topologies.

Our results have important implications for real systems. In the
social sciences they show the importance of opinion makers in
stabilizing the outcome of elections: weak external influences
result in an arbitrary but seemingly strong opinion that can switch
at random (see also \cite{redner2}), due to the arbitrary choice of
the ordered state in the weak perturbation regime. Thus, an elected
candidate winning a landslide election may have no solid support.
The slow dynamics can play a crucial role, since the time to
switching might occur only after the election day, especially if the
number of voters is large. Stronger external influences, counter
intuitively, reduce the relation time giving rise to improved
internal equilibration. In theoretical biology our results are
equivalent to the exact dynamical solution of the Wright-Fisher
model \cite{ewens} for arbitrary population sizes and mutation
rates. Our equations give not only the asymptotic equilibrium
distribution of alleles (see \cite{ewens} for approximate
expressions), but also its time evolution in two regimes, one where
mutations have difficulty overcoming an existing dominant allele
(the low perturbation regime) and one where random mutations
dominate (the high perturbation regime). Again this is crucial
information, since the equilibration time can be extremely long for
the typically small mutation rates observed in nature.

Finally, we emphasize that exact dynamical solutions, describing
systems out of equilibrium, are rare and important for the study of
many statistical properties that can be described in future publications.\\

M.A.M.A. and D.D.C. acknowledge financial support from CNPq and
FAPESP. I.R.E. was supported by National Science Foundation grant
CHE-0526866.


\end{document}